\UseRawInputEncoding
\documentclass[aps,prb,twocolumn,superscriptaddress,showpacs]{revtex4-1}
\usepackage{graphicx,amsmath,color,placeins}
\let\hide\iffalse
\usepackage[caption=false]{subfig}
\usepackage{epstopdf}

\begin{document}
\title{The exciton-polariton properties of hexagonal BN based microcavity and their potential applications in BEC and superconductivity}
\author{Huaiyuan Yang}
\affiliation{State Key Laboratory for Artificial Microstructure and Mesoscopic Physics, Frontier Science Center for Nano-optoelectronics and School of Physics, Peking University, Beijing 100871, P. R. China}
\author{Xinqiang Wang}
\affiliation{State Key Laboratory for Artificial Microstructure and Mesoscopic Physics, Frontier Science Center for Nano-optoelectronics and School of Physics, Peking University, Beijing 100871, P. R. China}
\affiliation{Collaborative Innovation Center of Quantum Matter, Peking University, Beijing 100871, P. R. China}
\author{Xin-Zheng Li}
\email{xzli@pku.edu.cn}
\affiliation{State Key Laboratory for Artificial Microstructure and Mesoscopic Physics, Frontier Science Center for Nano-optoelectronics and School of Physics, Peking University, Beijing 100871, P. R. China}
\affiliation{Interdisciplinary Institute of Light-Element Quantum Materials, Research Center for Light-Elemt Advanced Materials, and Collaborative Innovation Center of Quantum Matter, Peking University, Beijing 100871, People's Republic of China}
\affiliation{Peking University Yangtze Delta Institute of Optoelectronics, Nantong, Jiangsu 226010, People's Republic of China}
\date{\today}

\begin{abstract}
Microcavity exciton-polaritons are two-dimensional bosonic quasiparticles composed by excitons and photons.
Using model Hamiltonian with parameters generated from \textit{ab initio} density-functional theory and Bethe-Salpeter Equation calculations, we investigate the exciton and the exciton-polariton properties
of hexagnonal boron nitride (hBN) based microcavity.
We show that hBN based microcavities, including monolayer and all-dielectric ones, are promising
in optoelectronic applications.
Room temperature exciton-polariton Bose-Einstein Condensation can be achieved because of the large oscillator strength and binding energy of the exciton, and the strong interaction between
the exciton-polaritons and the longitudinal optical phonons.

Based on this BEC state, exciton-polariton mediated superconducting device can also be fabricated at
a few tens of Kelvin using the microcavity structure proposed by Laussy \textit{et al}.
\end{abstract}
\maketitle
\clearpage
	
\section{INTRODUCTION}
Excitons and photons can be strongly coupled in semiconductors, resulting in exciton-polariton~\cite{deng2010exciton}.
Microcavity is a platform to demonstrate the existence of this quasiparticle.
In these structure, excitons in the semiconductor quantum well (QW) couple with the
cavity photon~\cite{cavitypolaritons}.
The first experimental report of such exciton-polariton based on microcavity comes from
Weisbuch \textit{et al}, in a GaAlAs/GaAs QW system~\cite{PhysRevLett.69.3314}.
Formed by two kinds of bosons, exciton-polaritons are bosons by definition.
Under suitable environment, they can go through Bose-Einstein Condensation (BEC).
The lifetime of such microcavity exciton-polariton is typically short,
in order of several picoseconds.
Therefore, equilibrium BEC of exciton-polaritons in microcavity is normally hard to be
achieved in experiment.
By pumping excitons continuously into the system and with the help of polariton-polariton scattering and polariton-phonon scattering, this lifetime problem of the exciton-polariton can be circumvented.
One example of such dynamical nonequilibrium BEC state of exciton-polariton exists in CdTe-based microcavities~\cite{kasprzak2006bose}.
The extremely small effective mass of an exciton-polariton (typically $10^{-4}\sim10^{-5}m_0$ with $m_0$ being the free electron mass) means that such BEC state could occur at moderate temperatures if the Rabi coupling is large~\cite{plumhof2014room,Das2013Polariton,PhysRevLett.110.196406,2017Room}.
This remarkably increases the BEC temperature when compared to the conventional ultracold atomic systems (typically $\sim$nK).
After the establishment of exciton-polariton BEC, there are macroscopic occupations in the lower polariton branch (LP) $k=0$ state, and coherence exists.
This gives rise to possibility of fabricating high-quality laser without the requirement of population inversion~\cite{PhysRevA.53.4250, deng2003polariton}, which is probably one of the most
energy-consuming steps in conventional lasers.
Besides BEC, exciton-polariton is also relevant to the designing of superconducting devices.
In conventional Bardeen-Copper-Schrieffer (BCS) theory \cite{PhysRev.108.1175}, electron-phonon coupling plays an important role in mediating the attractive interactions between electrons.
Other mechanisms of superconductivity, however, also exist.
These mechanisms include the magnon mediated~\cite{PhysRevLett.117.076806, gong2017time}, photon mediated~\cite{PhysRevLett.15.524, PhysRevLett.122.133602}, the exciton mediated~\cite{PhysRevB.7.1020, Ginzburg:1976},
and the exciton-polariton mediated ones~\cite{PhysRevLett.104.106402,PhysRevLett.120.107001} etc.
By constructing a structure in which the BEC state of exciton-polariton can induce effective attractive interaction between electrons in the adjacent layer, Laussy \textit{et al.} proposed that the bogolons in exciton-polariton BEC could induce superconductivity~\cite{PhysRevLett.104.106402}.
Similar to the generation of the exciton-polariton based BEC state and the related fabrication of laser
as described in the former paragraph, creation of the exciton-polariton is the first step.
Large binding energy and strong oscillator strength of the exciton are favorable factors.
In the past decade, two-dimensional (2D) materials have received much attention due to their unique electronic and optical properties, applicable to nanoelectronics and nano-optical devices~\cite{wang2012electronics, xia2014two, balendhran2015elemental, RevModPhys.87.1139, kecik2018fundamentals, prete2020giant, 2015Exciton}.
Because of the reduced screening effects, the exciton binding energies are substantially larger
and the oscillator strength are normally stronger than their bulk correspondences.
This may result in more stable exciton-polaritons at moderate temperatures.
When the Rabi coupling is large, microcavity based on these 2D materials can provide
a better platform for the demonstration of the BEC state, the BEC-based laser, and the
exciton-polariton mediated superconductor than the conventional GaAs, GaN, and CdTe systems.
Among these 2D materials, hexagonal boron nitride (hBN) is unique due to its exceptionally
large exciton binding energies ($\sim$2~eV for monolayer and $\sim$0.7~eV for bulk)~\cite{PhysRevLett.96.126104, PhysRevLett.116.066803, PhysRevB.94.125303,PhysRevB.102.045117}.
The wide bandgap also means that when room temperature BEC is achieved, deep
ultraviolet laser is likely.
This is a feature highly desired in the lighting industry.
Based on such considerations, we explore in this manuscript the exciton-polariton related properties
in 2D materials based macrocavity, with a special emphasis on hBN.
The aim is to provide preliminary theoretical results, which could be helpful for
future experimental studies of exciton-polariton microcavities.
This paper is organized as follows.
In Sec.~II, we introduce the background theory for the methods we used in simulating these systems.

In Sec.~III, we modify and apply the theories to realistic problems. Then we present our results in detail.

These results include the properties of the exciton-polariton branches, the BEC state simulated by the
semiclassical Boltzmann equation, and the possibility for the superconducting microcavity structures.
In Sec.~IV, a brief summary of the conclusions is given.

\section{THEORY AND METHODS}
\subsection{Exciton-polaritons in microcavity}
In a microcavity(see Fig.1(a)(b) for a typical microcavity structure), photons are strongly confined in the $z$ direction. The energy dispersion of the
photon reads:
\begin{equation}
	E_{\mathrm{cav}}=\frac{\hbar c}{n} \sqrt{k_{z}^{2}+k_{\|}^{2}}.
\end{equation}
Here $n$ is the refractive index, $k_{\|}$ is the momentum of the photon in the $x-y$ plane, and $k_{z}=\frac{j \pi n}{L}$ ($j$=1, 2, 3...) is its discrete momentum in the $z$ direction.

The $j=1$ mode is the lowest one, and the other mode is well seperated in energy.
Therefore, we only consider $j=1$ mode throughout this manuscript.
The exciton energy dispersion is approximated by the effective mass model as:
\begin{equation}\label{eq2}
	E_{\text{exc}}=E_{\text{exc0}} + \frac{\hbar^{2}k_{\|}^{2}}{2m_{\text{exc}}},
\end{equation}
where $E_{\text{exc0}}$ is the exciton energy at the $\Gamma$ point, and $m_{\text{exc}}$ is the effective mass of the exciton to be fitted.
Under the rotating wave approximation, the exciton-polariton Hamiltonian can be described in a simple way as \cite{deng2010exciton}:
\begin{equation}
	\begin{aligned}
		\hat{H}_{\mathrm{pol}}=& \hat{H}_{\mathrm{cav}}+\hat{H}_{\mathrm{exc}}+\hat{H}_{I} \\
		=& \sum E_{\mathrm{cav}}\left(k_{\|}, k_{z}\right) \hat{a}_{k_{\|}}^{\dagger} \hat{a}_{k_{\|}}+\sum E_{\operatorname{exc}}\left(k_{\|}\right) \hat{b}_{k_{\|}}^{\dagger} \hat{b}_{k_{\|}} \\
		&+\sum g\left(\hat{a}_{k_{\|}}^{\dagger} \hat{b}_{k_{\|}}+\hat{a}_{k_{\|}} \hat{b}_{k_{\|}}^{\dagger}\right).
	\end{aligned}
\end{equation}
$g$ is the Rabi coupling, which is half of the Rabi splitting.

After the diagonalization, the polariton Hamiltonian becomes:
\begin{equation}
	\hat{H}_{\mathrm{pol}}=\sum E_{\mathrm{LP}} \hat{c}_{\text{LP},k_{\|}}^{\dagger} \hat{c}_{\text{LP},k_{\|}}
	+ \sum E_{\mathrm{UP}} \hat{c}_{\text{UP},k_{\|}}^{\dagger} \hat{c}_{\text{UP},k_{\|}}.
\end{equation}
Here the polariton operators are defined by:
\begin{equation}
 \hat{c}_{\text{LP/UP},k_{\|}} = C_{\text{LP/UP},k_{\|}} \hat{a}_{k_{\|}}
 + X_{\text{LP/UP},k_{\|}} \hat{b}_{k_{\|}}.
\end{equation}
$X$ and $C$ are the Hopfield coefficients obtained from the diagonalization.
The lower polariton (LP) and upper polariton (UP) energy dispersion equals:
\begin{equation}\label{polariton1}
	E_{\mathrm{LP}/ \mathrm{UP}}\left(k_{\|}\right)=\frac{1}{2}\left[E_{\mathrm{exc}}+E_{\mathrm{cav}} \pm \sqrt{4 g^{2}+\left(E_{\mathrm{exc}}-E_{\mathrm{cav}}\right)^{2}}\right].
\end{equation}
The lifetime of the LP mode ($\tau_{\text{pol}}$) is determined by\cite{deng2010exciton}:

\begin{equation}
	\frac{1}{\tau_{\text{pol}}} = \frac{|C|^{2}}{\tau_{\text{cav}}} + \frac{|X|^{2}}{\tau_{\text{exc}}}.
\end{equation}
They describe the factions for the component of the LP mode from the exciton and from the cavity photon.
$\tau_{\text{exc}}$ is the radiative lifetime of the exciton.
$\tau_{\text{cav}}$ is the lifetime of the cavity photon mode.

The effective mass of LP is given by\cite{deng2010exciton}:

\begin{equation}
	\frac{1}{m_{\mathrm{LP}}}=\frac{|C|^{2}}{m_{\mathrm{cav}}}+\frac{|X|^{2}}{m_{\mathrm{exc}}},
\end{equation}
where the cavity photon effective mass is $m_{\mathrm{cav}}=\frac{\hbar n j \pi}{c L} \sim 10^{-5}m_{0}$.
This leads to extremely small effective mass of the LP mode, which is crucial for the high
critical temperature of BEC based on this exciton-polariton.

\subsection{Excitonic properties from first principle}
Excitons are neutrally charged bosons formed by electron-hole pairs.
Bethe-Salpeter equation (BSE) is needed to describe them.
As many \textit{ab initio} softwares (such as YAMBO \cite{sangalli2019many} and BerkeleyGW \cite{DESLIPPE20121269}) are available, it is convenient to get the excitonic properties of a realistic material nowadays.
The BSE Hamiltonian reads:
\begin{equation}
	H_{\text{e} \text{e}^{\prime} \text{h} \text{h}^{\prime}}=\left(E_{\text{e}}-E_{\text{h}}\right) \delta_{\text{e} \text{h}, \text{e}^{\prime} \text{h}^{\prime}}+\left(f_{\text{e}}-f_{\text{h}}\right) \Xi_{\text{e} \text{e}^{\prime} \text{h} \text{h}^{\prime}},
\end{equation}
with $E_{\text{e(h)}}$ being the quasiparticle energy of the electron (hole) and $f_{\text{e(h)}}$ being
the occupation number.
$\Xi$ is the Bethe-Salpeter kernel calculated using the Kohn-Sham energies corrected by a scissor operator and Kohn-Sham orbitals, representing the interaction between an electron and a hole within the electron-hole pair.
After diagonalizing the BSE Hamiltonian, we get the exciton eigenenergy $E_{\lambda}$, with $\lambda$ labelling the excitonic states.
This determines the value of $E_{\text{exc0}}$ for each excitonic branch in Eq.~\ref{eq2}.
The difference between the bandgap $E_{\text{g}}$ and $E_{\lambda}$ is the exciton binding energy $E_{\text{b}}^{\lambda}$.
Following Ref.~\onlinecite{2015Exciton}, we calculate the exciton radiative lifetime from the BSE results.
In 2D materials, its value for an exciton in state $\lambda$ at zero temperature reads:
\begin{equation}
	\tau_{\mathrm{\lambda}}(0)=\frac{\hbar^{2} c}{8 \pi e^{2} E_{\mathrm{\lambda}}(0)} \frac{A_{\mathrm{uc}}}{\mu_{\mathrm{\lambda}}^{2}}.
\end{equation}
Here $\mu_{\mathrm{\lambda}}^{2}$ is the square modulus of the exciton transition dipole per unit cell
in the $x-y$ plane obtained from the BSE calculation, $A_{\mathrm{uc}}$ is the area of the unit cell, and $E_{\mathrm{\lambda}}(0)$ is the exciton energy with zero wavevector.
Using the effective mass approximation for the exciton dispersion~\cite{2015Exciton}, the average radiative lifetime in state $\lambda$ at temperature $T$ is:
\begin{equation}
	\tau_{\mathrm{\lambda}}^T=\tau_{\mathrm{\lambda}}(0) \frac{3}{4}\left(\frac{2 m^{\mathrm{\lambda}}_{\text{exc}} c^{2}}{E_{\mathrm{\lambda}}(0)^{2}}\right) k_{\mathrm{B}} T.
\end{equation}
$m^{\mathrm{\lambda}}_{\text{exc}}$ is the exciton effective mass and $k_{\mathrm{B}}$ is the Boltzmann constant.

\subsection{Exciton-polariton dispersion and Rabi coupling}
The exciton-polariton dispersion was not considered from first principle until recent years.
In 2015, Vasilevskiy \textit{et al.} presented a classical electrodynamics method to get the equation of motion for the exciton-polariton modes, using parameters generated by density-functional theory (DFT)
calculations~\cite{2015Ex}.

Parallel to this classical electrodynamics method, exciton-polariton dispersion in microcavity can also
be described by quantum mechanics, from quantized vector potential and exciton states\cite{1994Quantum, latini2019cavity}.
In this work, we follow the route of Savona \textit{et al.} in Ref.~\onlinecite{1994Quantum}, but use parameters
generated by first principle BSE calculations.

The differences between monolayer and bulk microcavity are addressed by taking limits for the width of the QW and that of the cavity.
Assuming a QW with a width of $L^{\prime}$ embedded in a cavity with a width of $L$, and the
excitons are uniformly confined, the equation of motion for the exciton-polariton reads (Eqs. 20-22 in Ref.~\onlinecite{1994Quantum}):
\begin{equation}\label{polariton2}
	\begin{array}{c}
		E^{2}-E_{\text{exc}}^{2}=\gamma \frac{E^{2}}{\alpha^{3}}[ k_{0} L^{\prime} \alpha+
		2 \cot \left(k_{0} \frac{L}{2} \alpha\right) \sin ^{2}\left(k_{0} \frac{L^{\prime}}{2} \alpha\right)
			\\-\sin \left(k_{0} L^{\prime} \alpha\right) ].
	\end{array}
\end{equation}
Here, $k_{0}=nE_{\text{exc}}/c$, and $\alpha=\sqrt{|E^{2}/E_{\text{exc}}^{2}-k_{\|}^{2}/k_{0}^{2}|}$.
$\gamma$ is the effective interaction constant, which is directly related to the exciton dipole moment $\mu_{\mathrm{\lambda}}$ by:
\begin{equation}
	\gamma= \frac{8 \pi|\mu_{\lambda}|^{2}}{\hbar n c A_{\text{uc}}\left(k_{0} L^{\prime}\right)^{2}}.
\end{equation}
$E$ appears on both sides of Eq.~\ref{polariton2}, meaning that it should be solved self-consistently
for each $k_{\|}$.
This leads to the energy dispersions of all the exciton-polariton branches.
For monolayer (bulk) microcavity, we choose the limit of $L^{\prime}$ approaching zero ($L^{\prime}=L$) and modify Eq.~\ref{polariton2} to get the dispersions in Sec.~\uppercase\expandafter{\romannumeral3}A\&B.
We find that the dispersions match perfertly with Eq.\ref{polariton1}.
By fitting with Eq.~\ref{polariton1}, the Rabi coupling $g$ can be
obtained.

\subsection{Exciton in screening environment}
Screening plays an important role in describing the excitonic properties.
The large binding energy in 2D materials comes from the reduced dielectric screening.
In real applications of optical devices, however, 2D materials are often sandwiched by dielectric materials.
The dielectric screening is expected to influence the excitonic properties.
It is much too computationally expensive to perform converged \textit{ab initio} BSE calculations with
this factor taken into account.
Therefore, we employed a 2D Wannier exciton model~\cite{wanniermodel}.
In this model, the exciton Schr\"odinger equation reads:
\begin{equation}\label{wanniermodel}
	\left[-\hbar^{2} \nabla_{r}^{2} / 2 \mu-e^{2} w^{2\text{D}}(r)\right] \psi_{\mathrm{exc}}(r)=E_{\text{b}} \psi_{\mathrm{exc}}(r).
\end{equation}
The screened Coulomb potential in two dimensions can be described by the Keldysh potential~\cite{keldysh1979coulomb}:
\begin{equation}\label{keldysh}
	w^{2\text{D}}(r)=\left[\mathcal{H}_{0}\left(\Sigma r / 2 r_{0}\right)-\mathcal{Y}_{0}\left(\Sigma r / 2 r_{0}\right)\right] / 8 \epsilon_{0} r_{0}.
\end{equation}
Here $\mathcal{H}_{0}$ ($\mathcal{Y}_{0}$) is the zeroth order Struve function (Bessel function of the second kind).
$\Sigma=\epsilon_{a}+\epsilon_{b}$ is the sum of the relative dielectric constants of the materials in both sides.
The screening length is $r_{0}=2\pi \alpha_{\text{2D}}$ and $\alpha_{\text{2D}}$ is the 2D polarizability:
\begin{equation}
	\alpha_{2 \text{D}}=\frac{2 e^{2}}{(2 \pi)^{2}} \sum_{c, v} \int_{\mathbf{k}} \frac{\left|\left\langle u_{c, \mathbf{k}}\left|\nabla_{\mathbf{k}}\right| u_{v, \mathbf{k}}\right\rangle\right|^{2}}{E_{c, \mathbf{k}}-E_{v, \mathbf{k}}} d^{2} \mathbf{k}.
\end{equation}
They can be obtained from first principle calculations.
After solving Eq.~\ref{wanniermodel}, we get the exciton binding energy $E_{\text{b}}$ and the exciton Bohr radius $a_{\text{B}}$ ($a_{\text{B}} = \int d^{2} \mathbf{r} \psi^{*}_{\text{exc}}(\mathbf{r}) r \psi_{\text{exc}}(\mathbf{r})$).

\subsection{Exciton-polariton BEC described by the semiclassical Boltzmann equation}	
We use the method of Porras \textit{et al.} to describe the nonequilibrium BEC in a semiclassical manner~\cite{2002Porras}.
The LP branch is divided into a lower polaritonic (lp) region and an exciton reservoir region.
The Boltzmann equation for the exciton-polariton population reads:
\begin{equation}\label{boltzmann}
	\begin{array}{c}
		\frac{d N_{k}^{\text{lp}}}{d t}=W_{k}^{\text{in}} n_{\text{x}}^{2}\left(1+N_{k}^{\text{lp}}\right)-W_{k}^{\text{out}} n_{\text{x}} N_{k}^{\text{lp}}-\Gamma_{k}^{\text{lp}} N_{k}^{\text{lp}} \\
\\
		\frac{d n_{\text{x}}}{d t}=-\frac{1}{S} \sum_{k} d g_{k}^{\text{lp}}\left[W_{k}^{\text{in}} n_{\text{x}}^{2}\left(1+N_{k}^{\text{lp}}\right)-W_{k}^{\text{out}} n_{\text{x}} N_{k}^{\text{lp}}\right] \\
		\quad-\Gamma_{x} n_{\text{x}}+p_{x},
	\end{array}
\end{equation}
where $N_{k}^{\text{lp}}$ and $n_{\text{x}}$ refer to the occupation number of the lower polariton region and the
density of the exciton reservoir.
$W_{k}^{\text{in}}$ and $W_{k}^{\text{out}}$ are the scattering rates.
$\Gamma_{k}^{\text{lp}}$ and $\Gamma_{\text{x}}$ are the loss rates.
Eq.~\ref{boltzmann} together with the equations describing the energy relaxation processes (details see Ref.~\onlinecite{2002Porras}) leads to a complete set of equations for the exciton-polariton dynamics.
After propagating the population, the reservoir density, and the reservoir energy density to their stationary solution, we can get the steady state for each pump strength.

\subsection{Exciton-polariton mediated superconductivity}
In conventional BCS theory, electron-phonon coupling induces effective attractive interaction between electrons.
Following this idea, Laussy \textit{et al.} proposed an exciton-polariton mediated mechanism for
superconductivity~\cite{PhysRevLett.104.106402}.
When BEC happens, the exciton-polariton density is tremendously large, the dipoles of the excitons can induce effective attractive interaction between the electrons in the neighboring region.
Here we investigate the superconductivity transition temperature of the neighboring electron gas layer
in this scenario.
We neglect the electron-phonon interaction and electron Coulomb interaction to highlight the exciton-polariton mechanism.
The microcavity structure is shown in Fig.~\ref{fig1}~(c).
The distance between the 2D electron gas QW and the exciton-polariton QW is $L$, which we set as several nanometers.
The reduced Hamiltonian after the Bogoliubov transformation and the mean-field approximation are made reads:
\begin{equation}
	\begin{array}{c}
		H=\sum_{\mathbf{k}} E_{\mathrm{el}}(\mathbf{k}) e_{\mathbf{k}}^{\dagger} e_{\mathbf{k}}+\sum_{\mathbf{k}} E_{\mathrm{bog}}(\mathbf{k}) b_{\mathbf{k}}^{\dagger} b_{\mathbf{k}}
		\\+\sum_{\mathbf{k}, \mathbf{q}} M(\mathbf{q}) e_{\mathbf{k}}^{\dagger} e_{\mathbf{k}+\mathbf{q}}\left(b_{-\mathbf{q}}^{\dagger}+b_{\mathbf{q}}\right).
	\end{array}
\end{equation}
$E_{\text{el}}(\mathbf{k})$ and $E_{\mathrm{bog}}(\mathbf{k})=\sqrt{\tilde{E}_{\mathrm{pol}}(\mathbf{k})\left(\tilde{E}_{\mathrm{pol}}(\mathbf{k})+2 U N_{0} A\right)}$
	($\tilde{E}_{\mathrm{pol}}(\mathbf{k}) \equiv E_{\mathrm{pol}}(\mathbf{k})-E_{\mathrm{pol}}(\mathbf{0})$)
are the in-plane dispersion of the electrons and the bogolons.
$N_{0}$ is the density of the condensed exciton-polaritons and $U=6a_{\text{B}}^{2}E_{\text{b}}X^{4}/A$ is the polariton-polariton interaction matrix element.
$A$ is the quantization area, $X$ is the exciton Hopfield coefficient.
The bogolon interacts with the electron through its excitonic fraction and the interaction strength is
\begin{equation}
	M(\mathbf{q})=\sqrt{N_{\text{C}} A} X V_{\mathrm{X}}(\mathbf{q}) \sqrt{\frac{E_{\mathrm{bog}}(\mathbf{q})-\tilde{E}_{\mathrm{pol}}(\mathbf{q})}{2 U N_{0} A-E_{\mathrm{bog}}(\mathbf{q})+\tilde{E}_{\mathrm{pol}}(\mathbf{q})}}\\ .
\end{equation}
$V_X$ is the electron-exciton interaction matrix element:
\begin{widetext}
\begin{equation}\label{vmatrix}
		V_{\mathrm{X}}(q)=\frac{e^{-q L}}{2 \epsilon A} \left\{\frac{e^2}{q\left[1+\left(\beta_{\text{e}} q a_{\text{B}} / 2\right)^{2}\right]^{3 / 2}}-\frac{e^2}{q\left[1+\left(\beta_{\text{h}} q a_{\text{B}} / 2\right)^{2}\right]^{3 / 2}}
		+\frac{ed\beta_{\text{e}}}{\left[1+\left(\beta_{\text{e}} q a_{\text{B}} / 2\right)^{2}\right]^{3 / 2}}+\frac{ed\beta_{\text{h}}}{\left[1+\left(\beta_{\text{h}} q a_{\text{B}} / 2\right)^{2}\right]^{3 / 2}}\right\},
\end{equation}
\end{widetext}
where $\beta_{\text{e(h)}}=m_{\text{e(h)}} /\left(m_{\text{e}}+m_{\text{h}}\right)$.
The first and second terms within the brace of Eq.~\ref{vmatrix} are negligible compared to the
third and forth dipolar ones.
The dipole $ed$ can be achieved by applying an electric field perpendicular to the 2D plane.
As the previous work in exciton-polariton mediated superconductivity ~\cite{PhysRevLett.104.106402,  laussy2012superconductivity, cherotchenko2016superconductivity}, the effective interaction between electrons can be expressed by Fr{\"o}hlich potential ~\cite{frohlich1952interaction}:
\begin{equation}\label{fp}
	V_{\mathrm{eff}}(\mathbf{q}, \omega)=\frac{2 M(\mathbf{q})^{2} E_{\mathrm{bog}}(\mathbf{q})}{(\hbar \omega)^{2}-E_{\mathrm{bog}}(\mathbf{q})^{2}}.
\end{equation}
Averaging the interaction over the 2D Fermi surface of the electrons, we get the effective
electron-electron interaction:
\begin{equation}\label{EFF}
	U_{0}(\omega)=\frac{A N(0)}{2 \pi} \int_{0}^{2 \pi}V_{\mathrm{eff}}(q, \omega) d \theta.
\end{equation}
Here $q$ equals $\sqrt{2k^2_F(1+\text{cos}\theta)}$, and $N(0)=m_{\text{e}} /\left(\pi \hbar^{2}\right)$ is the electron density of states at the Fermi surface.
$U_{0}$ is calculated numerically and substituted into the gap equation through
\begin{equation}\label{delta}
	\Delta(\xi, T)=-\int_{-\infty}^{+\infty} \frac{U_{0}\left(\xi-\xi^{\prime}\right) \Delta\left(\xi^{\prime}, T\right) \tanh \left(E / 2 k_{\mathrm{B}} T\right)}{2 E} d \xi^{\prime},
\end{equation}
where $E=\sqrt{\Delta\left(\xi^{\prime}, T\right) + \xi^{\prime 2}}$.
Eq.~\ref{delta} can be solved by iteration as long as the initial guess is rational.
The superconductivity can happen if the $\Delta(0, T)$ is non-zero.

These equations form the basis upon which superconductivity is discussed in this work.
We note, however, that they are only applicable in the adiabatic limit (the characteristic bogolon energy $\omega_\text{B}$ is much smaller than the Fermi energy $E_\text{F}$).
The final results presented in the manuscript is beyond this scenario.
To cope with this problem, the defects of the Fr{\"o}hlich potential will be discussed in Sec.~\uppercase\expandafter{\romannumeral3}.C.

\section{RESULTS}
\subsection{Monolayer hBN microcavity}
\begin{figure}[h]
	\includegraphics[width=1.0\linewidth]{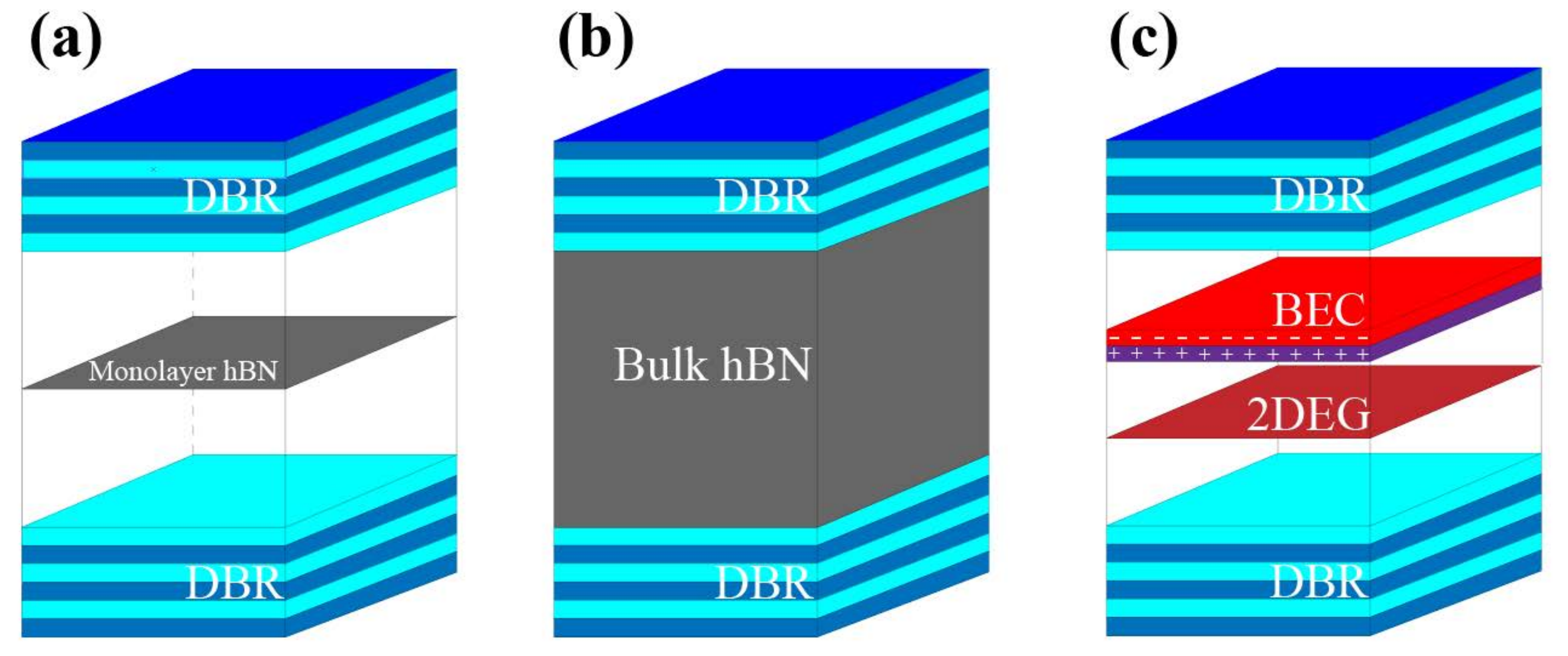}
	\caption{ \label{fig1}
		Sketch for (a) monolayer hBN microcavity, (b) all-dielectric hBN microcavity, and (c) model structure for the exciton-polariton mediated superconductivity.
		}
\end{figure}

The structure of the monolayer hBN based microcavity is shown in Fig.~\ref{fig1}~(a).
The monolayer is placed in the middle of the cavity and the dielectric materials fill the space between the distributed Bragg reflectors (DBRs).
In this cavity, $L^{\prime}$ in Eq.~\ref{polariton2} approaches 0, leading to $\frac{k_{0} L^{\prime} \alpha - \sin (k_{0} L^{\prime} \alpha)}{(k_{0} L^{\prime})^{2} } \rightarrow 0$ and $\frac{\sin^{2}(k_{0} \frac{L^{\prime}}{2} \alpha)}{(k_{0} L^{\prime})^{2} } \rightarrow \frac{\alpha^{2}}{4}$.
This means that Eq.~\ref{polariton2} becomes:
\begin{equation}\label{polariton3}
	\begin{array}{c}
		E^{2}-E_{\text{exc}}^{2}=\gamma^{\prime} \frac{E^{2}}{\alpha} \cot \left(k_{0} \frac{L}{2} \alpha\right),
	\end{array}
\end{equation}
where
\begin{equation}
	\gamma^{\prime}= \frac{4 \pi|\mu_{\lambda}|^{2}}{\hbar n c A_{\text{uc}}}.
\end{equation}
We apply Eq.~\ref{polariton3} to several typical 2D nitrides (BN, AlN, GaN and InN) and 2D transition metal dichalcogenides (TMDs, including MoS$_{2}$, MoSe$_{2}$, WS$_{2}$ and WSe$_{2}$), using existing \textit{ab initio} excitonic properties from BSE calculations~\cite{2020Giant, 2015Exciton}.
Table I shows the Rabi coupling of the lowest bright exciton in these materials when detuning is
zero and the refractive index is unity.
2D nitrides based microcavity have larger Rabi coupling than 2D TMDs in general, and hBN has the largest Rabi coupling among them.
This indicates that hBN is a very good candidate material for the exciton-polariton based properties to be investigated.

\begin{figure}[h]
	\includegraphics[width=1.0\linewidth]{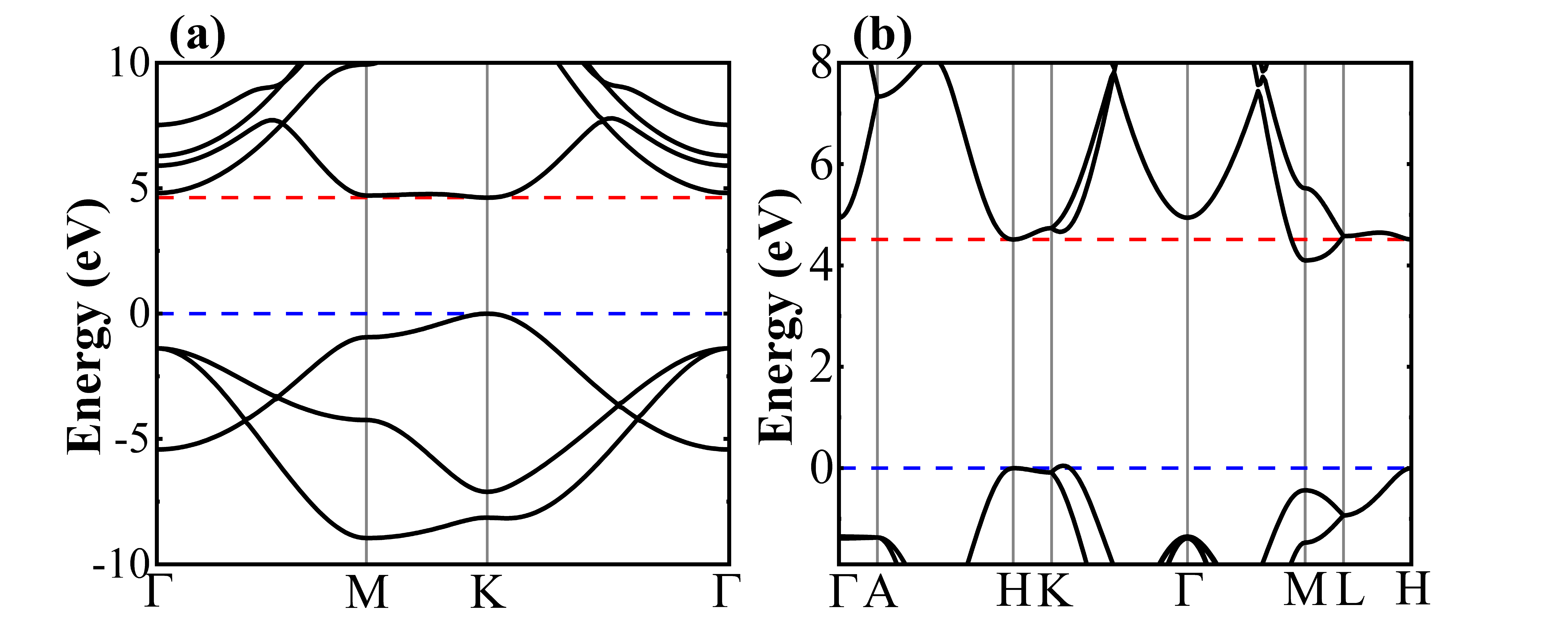}
	\caption{ \label{fig2}
		LDA band structure of (a) monolayer hBN and (b) bulk hBN microcavity.
		The distance between the red and blue dashed line indicate the direct bandgap.
}
\end{figure}

\begin{figure}[h]
	\includegraphics[width=1.0\linewidth]{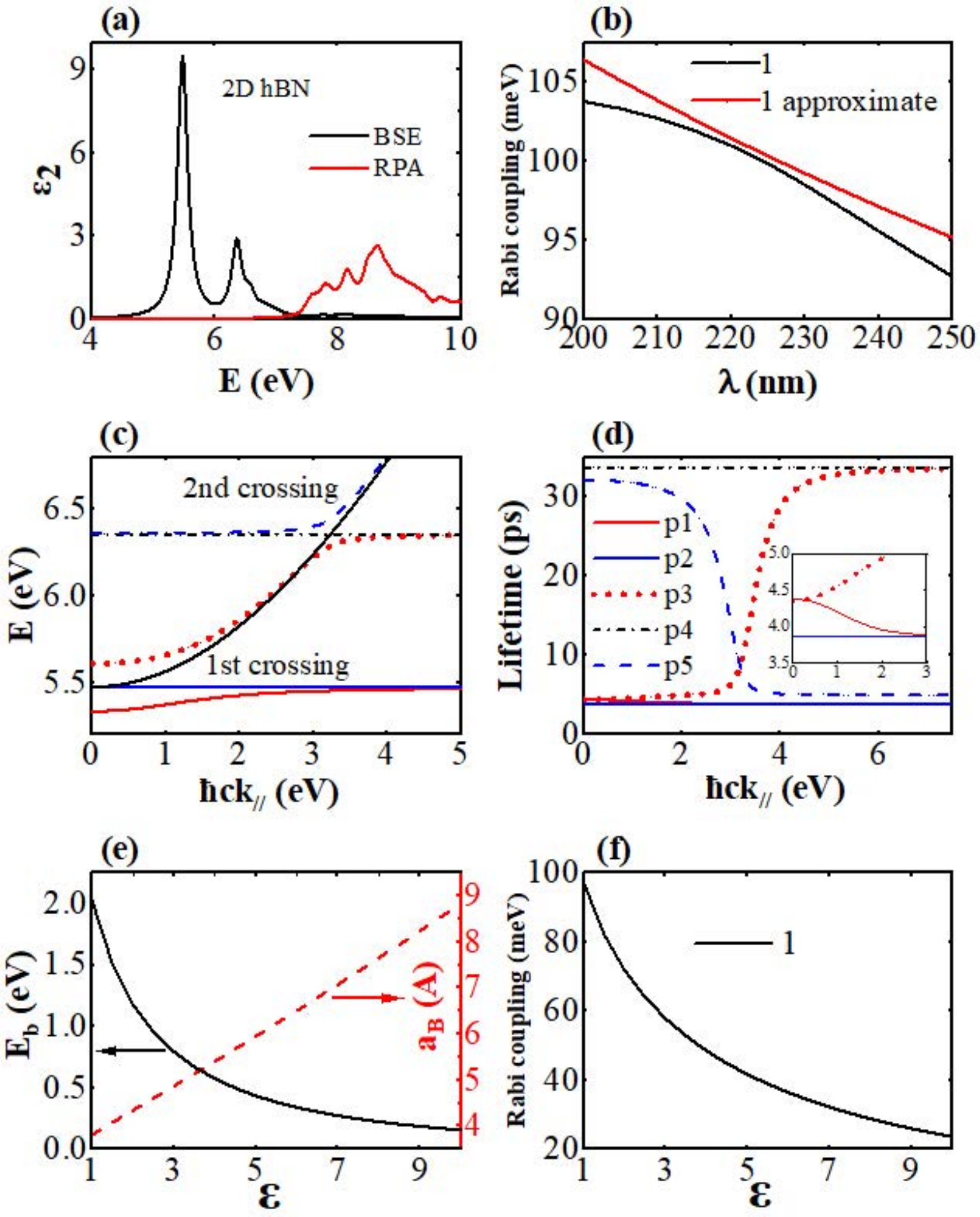}
	\caption{ \label{fig3}
		Exciton and polariton properties for monolayer hBN micrcavity.
    	(a) Absorption spectrum of 2D hBN calculated by BSE (black line) and RPA (red line).
		(b) Rabi coupling of polaritons composited by cavity photon mode and exciton 1(black line) for different photon wavelength using Eq.~\ref{polariton3}.
		The corresponding results using approximate expression Eq.~\ref{approx} are in red line.
		(c) Dispersion of the five branches of polaritons(labels are in (d)) and cavity photon (black solid line) when zero detuning.
		(d) Polariton lifetimes for the five branches of polaritons.
		(e) Binding energy(black solid line) and radius of the exciton(red dashed line) under different dielectric screening environment.
		(f) Rabi coupling of polaritons composited by cavity photon mode and exciton 1 under different dielectric screening environment.
}
\end{figure}

\begin{table}[htb]
	\centering
	\caption{
		Rabi coupling of the lowest bright exciton in eight typical 2D TMDs and 2D nitrides calculated by Eq.~\ref{polariton3} based on existing BSE results from Refs.~\onlinecite{2020Giant} and~\onlinecite{2015Exciton}. Detuning is set to zero, and the refractive index $n$ is unity.
The unit of Rabi coupling $g$ is meV.
	}
	\begin{tabular}{cc|cc}
	\hline
	\hline
	\textrm{Materials}&
	\textrm{$g$}&
	\textrm{Materials}&
	\textrm{$g$}\\
	\hline
	MoS$_{2}$ & 20.8 &  BN & 99.4\\
	MoSe$_{2}$ & 19.0 & AlN & 85.4\\
	WS$_{2}$ & 23.5 & GaN & 70.3\\
	WSe$_{2}$ & 20.1 & InN & 32.4\\
	\hline
	\hline
	\end{tabular}
\end{table}
For a more detailed understanding of the excitonic properties of hBN, we performed DFT calculations
using QUANTUM ESPRESSO (QE) and then the BSE calculations using YAMBO~\cite{giannozzi2009quantum, 2009yambo, 2019Many}.
This subsection focusses on the monolayer.
Local density approximation (LDA) is used in describing the Kohn-Sham exchange-correlation
potential, along with a $16\times16\times1$ k-point mesh for Brillouin-zone sampling and a kinetic energy
cutoff of 80 Ry for the expansion of the wavefunctions.
The direct LDA bandgap is 4.62 eV at $K$ as shown in Fig.~\ref{fig2}~(a).
In the BSE calculation, a scissor operator of 2.87~eV is applied to the Kohn-Sham energies, and
a denser k-point mesh of $36\times36\times1$ is used to converge the results.
The transitions between all four valence bands and the lowest four conduction bands are included.
An exciton binding energy of 2.02~eV is obtained.
The lowest bright exciton is mainly contributed by the transition around the $K$ point.
Optical absorption spectrum at the random-phase approximation (RPA) level and at the BSE level are
shown in Fig.~\ref{fig3}~(a).
The binding energy and the absorption spectrum are similar to recent studies~\cite{ferreira2019excitons, 2020Giant}.
Compared to the RPA result, the BSE spectrum captures the excitonic effects, which substantially redshift the optical peak and increase the intensity of the absorption edge.
The four lowest exciton states (labeled as 1, 2, 3, and 4) have prominent oscillator strengthes (larger than $10\%$ of the maximum value).
The degenerate exciton states 1 and 2 (3 and 4) are 1s (2p) states coming from $K$ and $K'$ respectively.
Their detailed information, including the positions of the absorption energy, the effective interaction constant, and the radiative lifetimes at zero and room temperatures, are shown in Table II.
Due to large oscillator strength, the zero temperature radiative lifetimes of these four exciton
states are very short, of the $10^1\sim10^2$ fs.
Combined with the exciton effective mass extracted from the \textit{ab initio} calculations~\cite{PhysRevB.98.125206}, we calculate the room temperature radiative lifetimes in the last column.
They are at the order of $10^1$ ps.
The increase of the radiative lifetime at finite temperature is mainly due to the thermal distribution of the excitonic states, away from the edge.
\begin{table}[htb]
	\centering
	\caption{
		Exciton energy, effective interaction constant, Rabi coupling at zero detuning, radiative lifetimes at 0~K ($\tau_{\lambda}(0)$) and average radiative lifetimes at room temperature ($\tau_{\lambda}^{\text{RT}}$) for the 1, 2, 3, and 4 exciton states in 2D hBN.
	}
	\begin{tabular}{cccccc}
		\hline
		\hline
		\textrm{Index}&
		\textrm{E(eV)}&
		\textrm{$\gamma^{\prime}$(a.u.)}&
		\textrm{$g$(meV)}&
		\textrm{$\tau_{\lambda}(0)$(fs)}&
		\textrm{$\tau_{\lambda}^{\text{RT}}$(ps)}\\
		\hline
		1 & 5.473 & 5.486 &  99.43& 29.15 & 30.85\\
		2 & 5.473 & 5.486 &  99.43& 29.15 & 2.06\\
		3 & 6.356 & 1.561 &  53.04& 88.22 & 76.60\\
		4 & 6.356 & 1.561 &  53.04& 88.22 & 21.62\\
		\hline
		\hline
	\end{tabular}
\end{table}

Rabi coupling at different photon wavelength of the lowest two exciton is shown in Fig.~\ref{fig3}~(b)
when $n=1$.
Its magnitude is large, which decreases only by $\sim$10~meV as the photon wavelength increases.

\begin{figure}[h]
	\includegraphics[width=1\linewidth]{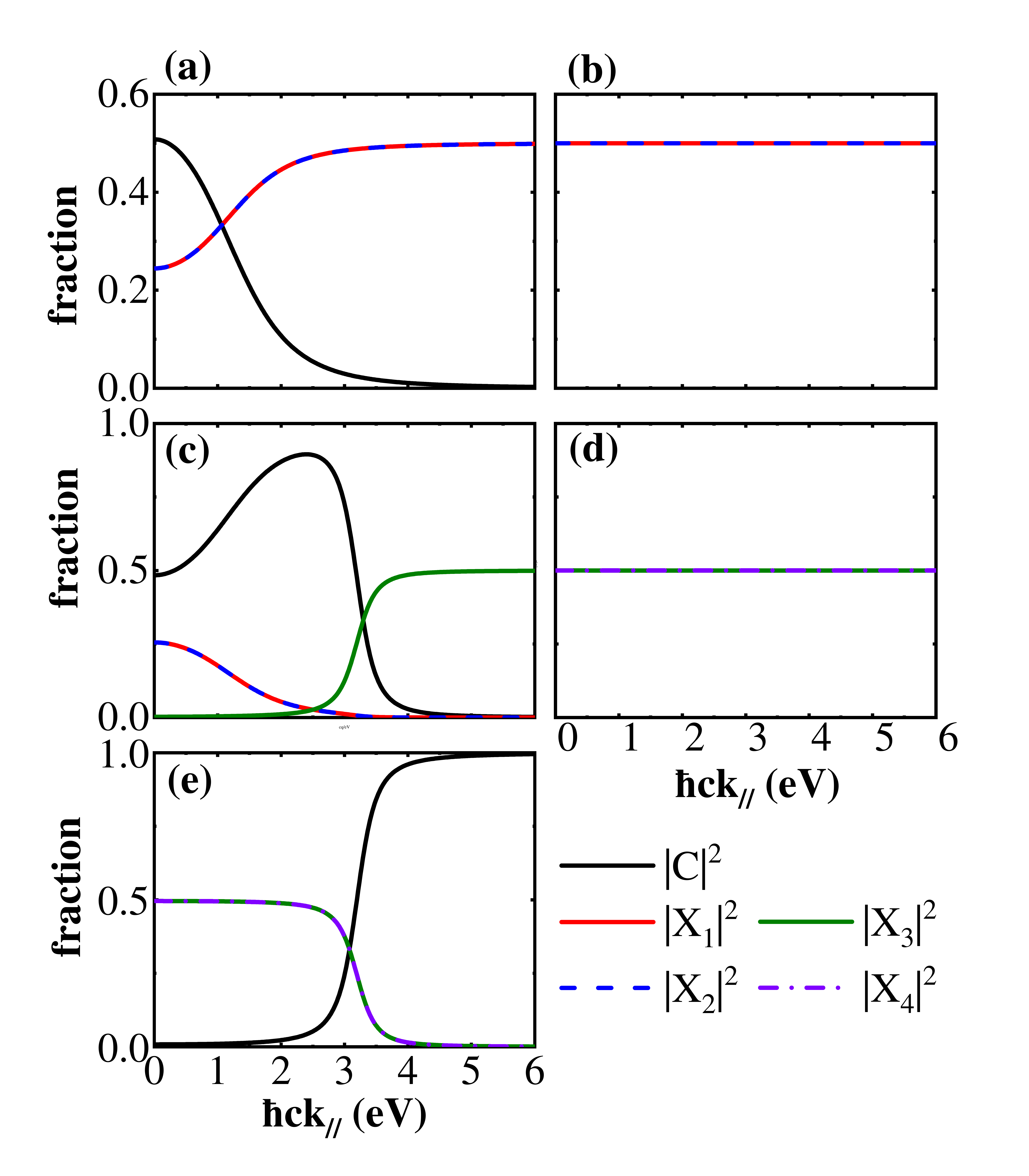}
	\caption{ \label{fig4}
		(a)-(e)Non-zero Hopfield coefficients for monolayer hBN microcavity polaritons at zero detuning for the first polariton mode to the fifth polariton mode.
	}
\end{figure}

The approximate expression for the Rabi coupling~\cite{savona1997optical, sarchi2007bose},
\begin{equation}\label{approx}
	\begin{array}{c}
		\hbar g_{\lambda} \simeq  \sqrt{\frac{2 \pi E_{\lambda}^{\text{exc}}\mu_{\lambda}^{2}}{n^{2}LA_{\text{uc}}}},
	\end{array}
\end{equation}
also matches well with the results of Eq.~\ref{polariton3}.
To quantify the coupling between the four excitonic states and the cavity photon mode, we construct a Hamiltonian:
\begin{equation}
	\hat{H}_{\mathrm{pol}}=
	\left(\begin{array}{ccccc}
		E_{\text{cav}} & g_{1} & g_{2} & g_{3} & g_{4} \\
		g_{1} & E_{\text{exc1}} & 0 & 0 & 0\\
		g_{2} & 0 & E_{\text{exc2}} & 0 & 0\\
		g_{3} & 0 & 0 & E_{\text{exc3}} & 0\\
		g_{4} & 0 & 0 & 0 & E_{\text{exc4}} \\
	\end{array}\right).
\end{equation}
Diagonalizing this Hamiltonian, we get five exciton-polariton modes (labeled as p1, p2, p3, p4 and p5, shown in Fig.~\ref{fig3}~(c)).
As $E_{\text{exc1}}=E_{\text{exc2}}$, $E_{\text{exc3}}=E_{\text{exc4}}$ and $E_{\text{exc3}}-E_{\text{exc1}} \gg g$, the splitting between p1 and p3 (p3 and p5) equals $2\sqrt{g_{1}^2 + g_{2}^2}$ ($2\sqrt{g_{3}^2 + g_{4}^2}$).
Each of the five exciton-polariton modes is a linear combination of the original four excitonic states and one cavity photon mode.
The corresponding Hopfield coefficients are labelled as $C$ for the photon mode and $X_1$, $X_2$, $X_3$, $X_4$ for the excitonic ones.
In Fig.~\ref{fig4}, we show the square modulus of these coefficients.
As can be seen in Fig.~\ref{fig4}~(b), the p2 exciton-polariton mode is comprised by the
contributions from $X_1$ and $X_2$, meaning that it is purely excitonic state.
The same feature holds for the p4 exciton-polariton mode as it is comprised purely by
contributions from $X_3$ and $X_4$ (Fig.~\ref{fig4}~(d)).
The p1, p3, and p5 exciton-polariton modes, on the other hand, show strong features of coupling,
especially in the region when the dispersion of the cavity photon mode intersects with the
exciton energies.
This is clearly seen if we compare the dispersion of the exciton-polariton modes in
Fig.~\ref{fig3}~(c) with the analysis of the Hopfield coefficients in Fig.~\ref{fig4}.
As $E_{\text{exc3}}-E_{\text{exc1}} \gg g$, p1 is composed by the photon mode and the
first degenerate exciton pair before the crossing near $E_{\text{exc1}}$(1st crossing in Fig.~\ref{fig3}~(c)) and it is
a bare excitonic state after that.
p5 is dominated by the second degenerate exciton pair and photon fraction before and after
the crossing near $E_{\text{exc3}}$(2nd crossing in Fig.~\ref{fig3}~(c)).
p3 is a mixture of all five components as it goes through both coupling regions.
It is composed purely by the second degenerate exciton pair after the 2nd crossing.
The lifetimes of these five exciton-polariton modes are evaluated by:
\begin{equation}
	\begin{array}{c}
		\frac{1}{\tau_{\text{pol}}} = \frac{|C|^{2}}{\tau_{\text{cav}}} + \frac{|X_{1}|^{2}}{\tau_{\text{exc1}}} + \frac{|X_{2}|^{2}}{\tau_{\text{exc2}}} +
		\frac{|X_{3}|^{2}}{\tau_{\text{exc3}}} + \frac{|X_{4}|^{2}}{\tau_{\text{exc4}}}.
	\end{array}
\end{equation}
We assume a typical cavity photon lifetime of 5~ps for high quality microcavity.
The results are shown in Fig.~\ref{fig3}~(d), which can also be understood by analyzing the composition
of each exciton-polariton mode in Fig.~\ref{fig4}.
The lifetimes of the p2 (p4) exciton-polariton mode is 3.6~ps (31.6~ps).
They are constant with respect to $k_\parallel$ in Fig.~\ref{fig3}~(d), as they compositions
donot change with $k_\parallel$ in Fig.~\ref{fig4}.
As the lifetimes of the exciton state 2 in
 and the photon mode (5~ps) are short compared
with the others, the exciton-polariton modes with large fraction of them have short lifetimes.
This leads to short lifetimes of the p1, p2, p3 modes before the 2nd crossing
and that of the p5 mode after it.

In reality, the dielectric materials filling the empty space in Fig.1(a) induce screening.
Here we quantify the screening effects on the exciton binding energy and Rabi coupling using Eqs.~\ref{wanniermodel} and~\ref{keldysh}.
The screening length $r_0$ is adjusted to match the \textit{ab initio} binding energy in vacuum.
The resulting binding energy and the exciton Bohr radius as a function of the dielectric constant
is shown in Fig.~\ref{fig3}~(e).
The binding energy is substantially reduced by the environmental screening and the radius increases
linearly with the dielectric constant.
Under moderate temperatures, the exciton is still stable with large dielectric constant.
Based on such excitonic properties after screening, we evaluate the Rabi coupling using Eq.~\ref{polariton3}.
The exciton oscillator strength can be approximated by:
\begin{equation}
		|\mu_{\lambda}|^{2} \propto \frac{\left|p_{0}\right|^{2}\left|\psi_{\lambda}(0)\right|^{2}}{E_{\lambda}}.
\end{equation}
Here $p_{0}$ is the coupling strength, which we assume as constant.
$E_\lambda$ is the exciton energy, which equals $E_{\text{g}}-E_{\text{b}}$.
$E_{\text{g}}$ is approximated by $E_{\text{g}}=1.14E_{\text{b}}+5.07$, which is fitted from \textit{ab initio} results\cite{Guo_2021}.
$\psi_{\lambda}(0)$ is the exciton wavefunction at zero relative distance between the electron and
the hole within the pair.
The Rabi coupling decreases with increasing dielectric constant as shown in Fig.~\ref{fig3}~(f).
Under moderate dielectric screening, it is still larger
than the room temperature thermal energy, indicating the stability the exciton-polaritons.

The most attractive property of such exciton-polariton is the possibility of room temperature BEC.
Using semiclassical Boltzmann equation, plenty of works have been reported on the possibility and mechanism that drive the microcavity exciton-polaritons to the lowest state ($k_{\|}$ state in the LP) to achieve BEC~\cite{1997Bottleneck, 2002Polariton, 2002Porras,2005Condensation, 2010Numerical}.
It has been proven that polariton-polariton interaction and polariton-phonon interaction are essential in the relaxation process.
In conventional semiconductors such as GaAs or CdTe, the Rabi coupling is so small that the optical phonons cannot participate in this process and only acoustic phonons contribute~\cite{1997Bottleneck, 2002Polariton,2010Numerical}.
For the relaxation process of exciton-polariton in 2D hBN microcavity, scattering with accoustic phonon is negligible as acoustic phonon dispersions are too flat to fulfill the energy and momentum conservation law.
Optical phonons, on the other hand, participate in this process.
To theoretically describe this process, we follow a two-step strategy.
The interaction between exciton-polaritons is derived first by taking into account of the interaction between their original components.
The Hamiltonian is first written as:
\begin{equation}
		\begin{array}{c}
			H_{\text{I}}=\sum_{\mathbf{k}_{1}, \mathbf{k}_{2}, \mathbf{q}} \frac{M_{\text{xx}}}{2 S} (a_{\mathbf{k}_{1}}^{\dagger} a_{\mathbf{k}_{2}}^{\dagger} a_{\mathbf{k}_{1}+\mathbf{q}} a_{\mathbf{k}_{2}-\mathbf{q}} + b_{\mathbf{k}_{1}}^{\dagger} b_{\mathbf{k}_{2}}^{\dagger} b_{\mathbf{k}_{1}+\mathbf{q}} b_{\mathbf{k}_{2}-\mathbf{q}}) \\+
			\frac{\sigma_{\text{sat}}}{ S}(a_{\mathbf{k}_{1}}^{\dagger} a_{\mathbf{k}_{2}}^{\dagger} a_{\mathbf{k}_{1}+\mathbf{q}} c_{\mathbf{k}_{2}-\mathbf{q}}+ b_{\mathbf{k}_{1}}^{\dagger} b_{\mathbf{k}_{2}}^{\dagger} b_{\mathbf{k}_{1}+\mathbf{q}} c_{\mathbf{k}_{2}-\mathbf{q}}) + \text{H.C.}
		\end{array}
\end{equation}
We only consider p1, p2 and p3 exciton-polaritons.
$a$ and $b$ are the operators for the 1 and 2 excitonic states, and $c$ is the cavity photon operator.
$M_{\text{xx}}=6E_{\text{b}}a_{\text{B}}^2$ and $\sigma_{\text{sat}}=3.6ga_{\text{B}}^2$ are exciton-exciton and exciton-photon interaction strengthes respectively.
Transforming $a$, $b$ and $c$ into the exciton-polariton basis:
	\begin{equation}
		\begin{array}{l}
			p_{\text{p}1, \mathbf{k}}=X^{\text{p}1}_{1, \mathbf{k}} a_{\mathbf{k}} + X^{\text{p}1}_{2, \mathbf{k}} b_{\mathbf{k}} + C^{\text{p}1}_{\mathbf{k}} c_{\mathbf{k}}, \\
			p_{\text{p}2, \mathbf{k}}=X^{\text{p}2}_{1, \mathbf{k}} a_{\mathbf{k}} + X^{\text{p}2}_{2, \mathbf{k}} b_{\mathbf{k}} + C^{\text{p}2}_{\mathbf{k}} c_{\mathbf{k}}, \\
			p_{\text{p}3, \mathbf{k}}=X^{\text{p}3}_{1, \mathbf{k}} a_{\mathbf{k}} + X^{\text{p}3}_{2, \mathbf{k}} b_{\mathbf{k}} + C^{\text{p}3}_{\mathbf{k}} c_{\mathbf{k}},
		\end{array}
	\end{equation}
representing $H_{\text{I}}$ in terms of exciton-polariton operators and neglecting the p2 and p3 branch, we can get the lowest polariton branch:
\begin{equation}
	\begin{array}{c}
			H_{\text{p}1}=\sum_{\mathbf{k}} \epsilon_{\text{p}1,\mathbf{k}} p_{\text{p}1, \mathbf{k}}^{\dagger} p_{\text{p}1, \mathbf{k}} \\+
			\sum_{\mathbf{k}_{1}, \mathbf{k}_{2}, \mathbf{k}_{3}, \mathbf{k}_{4}} \frac{V^{\text{p}1-\text{p}1}_{\mathbf{k}_{1}, \mathbf{k}_{2}, \mathbf{k}_{3}, \mathbf{k}_{4}}}{ 2S}p_{\text{p}1, \mathbf{k_{1}}}^{\dagger} p_{\text{p}1, \mathbf{k_{2}}}^{\dagger} p_{\text{p}1, \mathbf{k_{3}}} p_{\text{p}1, \mathbf{k_{4}}},
	\end{array}
\end{equation}
where
\begin{equation}
	\begin{array}{c}
			\frac{1}{2} V_{\mathbf{k}_{1}, \mathbf{k}_{2}, \mathbf{k}_{3}, \mathbf{k}_{4}}^{\text{p}1- \text{p}1}=(\frac{1}{2} M_{\text{x x}} X^{\text{p}1}_{1, \mathbf{k}_{1}} X^{\text{p}1}_{1, \mathbf{k}_{2}} X^{\text{p}1}_{1, \mathbf{k}_{3}} X^{\text{p}1}_{1, \mathbf{k}_{4}}
			\\+ \frac{1}{2} M_{\text{x x}} X^{\text{p}1}_{2, \mathbf{k}_{1}} X^{\text{p}1}_{2, \mathbf{k}_{2}} X^{\text{p}1}_{2, \mathbf{k}_{3}} X^{\text{p}1}_{2, \mathbf{k}_{4}}
			+ \sigma_{\text {sat}} C_{\mathbf{k}_{1}}^{ \text{p}1} X_{1, \mathbf{k}_{2}}^{\text{p}1} X_{1, \mathbf{k}_{3}}^{\text{p}1} X_{1, \mathbf{k}_{4}}^{\text{p}1}
			\\+ \sigma_{\text{sat}} X_{1, \mathbf{k}_{1}}^{\text{p}1} X_{1, \mathbf{k}_{1}}^{\text{p}1} C_{\mathbf{k}_{3}}^{\text{p}1} X_{1, \mathbf{k}_{4}}^{\text{p}1}
			+ \sigma_{\text {sat}} C_{\mathbf{k}_{1}}^{\text{p}1} X_{2, \mathbf{k}_{2}}^{\text{p}1} X_{2, \mathbf{k}_{3}}^{\text{p}1} X_{2, \mathbf{k}_{4}}^{\text{p}1} \\
			+ \sigma_{\text{sat}} X_{2, \mathbf{k}_{1}}^{\text{p}1} X_{2, \mathbf{k}_{1}}^{\text{p}1} C_{\mathbf{k}_{3}}^{\text{p}1} X_{2, \mathbf{k}_{4}}^{\text{p}1} ) \delta_{\mathbf{k}_{1}+\mathbf{k}_{2}, \mathbf{k}_{3}+\mathbf{k}_{4}}.
	\end{array}
\end{equation}	

The coupling between the exciton, photon and phonon has received much attention, and an elementary excitation named ``phonoriton'' was proposed to describe a quasiparticle resulting from their interactions~\cite{latini2021phonoritons}.
We note, however, that the coupling between the exciton and photon is much stronger in microcavity
when the Rabi coupling is large.
Here we carry out our simulations in this scenario and investigate the exciton-polariton relaxation
process toward BEC.
As will be demonstrated, the scattering between exciton-polariton and phonon plays an important role.
The interactions between the LO phonons and the exciton-polariton were taken into account through:
\begin{equation}
	\begin{aligned}
		H_{\mathrm{p1}-\mathrm{ph}}=& \sum_{\mathbf{k}, \mathbf{q}} M(|\vec{k}, \vec{q}|) \times\left(c_{\text{LO},\mathbf{q}}-c_{\text{LO},-\mathbf{q},}^{\dagger}\right) p_{\mathrm{p1},\mathbf{k}+\mathbf{q}}^{\dagger} p_{\mathrm{p1},\mathbf{k}}.
	\end{aligned}
\end{equation}
The interaction matrix element can be described by the Fr{\"o}hlich model \cite{frol1,frol2,2010Numerical,PhysRevB.103.235424} as:
\begin{equation}\label{Fint}
	\begin{array}{c}
			M(\mathbf{k}, \mathbf{q})=i X_{\mathbf{k}}X_{ \mathbf{k^{\prime}}} \sqrt{\frac{2 \pi e^{2} \hbar \omega_{\text{LO}}}{\vec{q}_{\|}^{2}V}\left(\frac{1}{\epsilon_{\infty}}-\frac{1}{\epsilon_{0}}\right)}
			\left[ I_{\text{e}}^{\|}(|\mathbf{q}|) - I_{\text{h}}^{\|}(|\mathbf{q}|) \right],
	\end{array}
\end{equation}
where
\begin{equation}
		I_{\text{e(h)}}^{\|}=\left[1+\left(\frac{m_{\text{e(h)}}}{2 m_{\text{exc}}}\left|q_{\|}\right| a_{\text{B}}\right)^{2}\right]^{-3 / 2},
\end{equation}
and $X_{\mathbf{k}}$ and $X_{ \mathbf{k^{\prime}}}$ are Hopfield coefficients.

\begin{figure}[h]
			\includegraphics[width=1.0\linewidth]{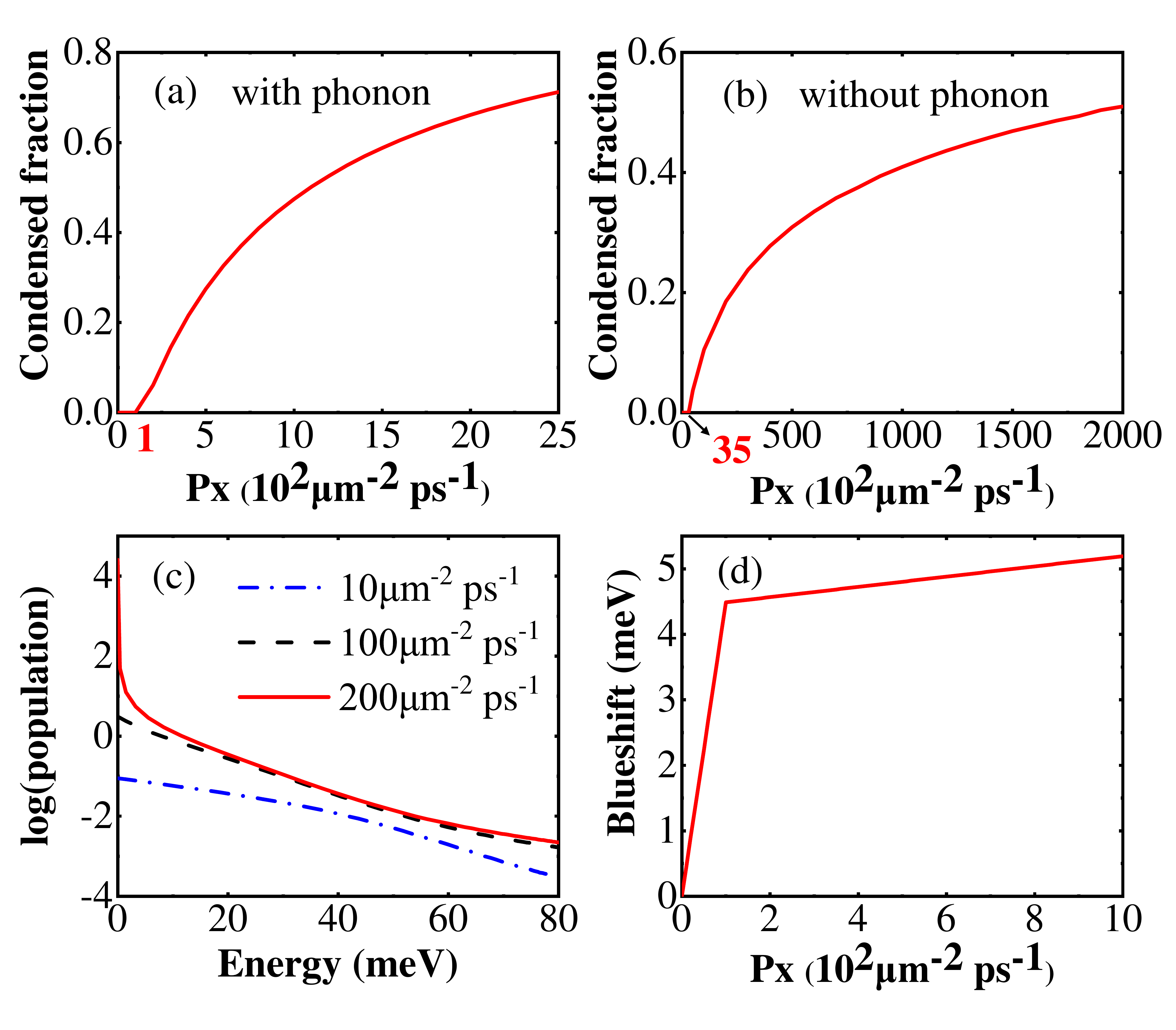}
			\caption{ \label{fig5}
				Results of room temperature BEC analysis using Boltzmann equation and generalized GP equation.
				The cavity photon wavelength is 230nm.
				(a) Condensed fraction when the optical phonons are included.
				(b) Condensed fraction without optical phonons.
				(c) Three typical population curve under three different pump strength.
				(d) Energy blueshift of lowest polariton mode under different pump strength.
			}
\end{figure}

After these treatments, we solve the Boltzmann equation using the method described in Section.II.E with polariton-polariton scattering and polariton-LO phonon scattering included.
The results are shown in Fig.~\ref{fig5}(a) and (b).
The pumping threshold is $\rm 1\times 10^2 \mu m^{-2} ps^{-2}$ ($\rm 35\times 10^2 \mu m^{-2} ps^{-2}$) respectively when polariton-LO phonon scattering is (is not) considered.

Although polariton-polariton scattering alone is enough to achieve BEC, polariton-LO phonon scattering plays an important role in the exciton-polariton relaxation process.
It lowers the pumping threshold by nearly two orders of magnitude.
The strong polariton-LO phonon scattering comes from the large Fr{\"o}hlich interaction matrix element in Eq.\ref{Fint}.
It is large when the transfer momentum $q$ is small, which is the case in the exciton-polariton relaxation process.
The population in logarithm scale is shown in Fig.~\ref{fig5}~(c).
As the pumping strength is higher, the population changes to Boltzmann distribution (black line) at threshold, and to Bose-Einstein distribution (red line) when the pumping strength is higher than the threshold.
One side effect for the use of pumping is that the energy of the exciton-polariton mode will go through a renormalization.
We can estimate this using the generalized Gross-Pitaevskii (GP) equation \cite{PhysRevLett.99.140402,2017Room}:
\begin{equation}
	\begin{array}{l}
			i \hbar \frac{\mathrm{d} \psi}{\mathrm{d} t}=\left[E_{0}+\alpha|\psi|^{2}+g_{\mathrm{R}} n_{\mathrm{R}}+\frac{i}{2}\left(\hbar r n_{\mathrm{R}}-\Gamma\right)\right] \psi, \\
			\frac{\mathrm{d} n_{\mathrm{R}}}{\mathrm{d} t}=P-\left(\Gamma_{\mathrm{R}}+r|\psi|^{2}\right) n_{\mathrm{R}}.
	\end{array}
\end{equation}
Here $\psi$ and $n_{\mathrm{R}}$ are the exciton-polariton field and the reservoir density.
$E_{0}$ is the bare exciton-polariton energy.
$\alpha$ and $g_{\mathrm{R}}$ are the polariton-polariton interaction and lower polariton-reservoir exciton interaction strengthes.
$\Gamma$ and $\Gamma_{\text{R}}$ are the exciton-polariton decay rates.
$r$ is the condensation rate (including contributions from both polariton-polariton and polariton-phonon scattering), and $P$ is the pumping rate.
All of these parameters can be estimated from each specific material.
The threshold pumping rate is $P_{0}=\frac{\Gamma \Gamma_{\text{R}}}{\hbar r}$.
The exciton-polariton energy shift above and below the threshold is evaluated by:
\begin{equation}
		\begin{array}{l}
				\Delta E_{\text{below}} =\frac{g_{\text{R}}P}{\Gamma_{\text{R}}},\\
				\Delta E_{\text{above}} = \frac{\hbar P \alpha}{\Gamma} - \frac{\Gamma_{\text{R}}\alpha}{r} +\frac{g_{\text{R}}\Gamma}{\hbar r}.
		\end{array}
\end{equation}
The result for the 2D hBN microcavity is shown in Fig.~\ref{fig5}~(d).
The slope is smaller above the threshold than before.
The blueshift is of the meV order.
These values are small and can be reasonably neglected in the Boltzmann equation.

\subsection{All-dielectric hBN microcavity}
\begin{figure}[h]
		\includegraphics[width=1.0\linewidth]{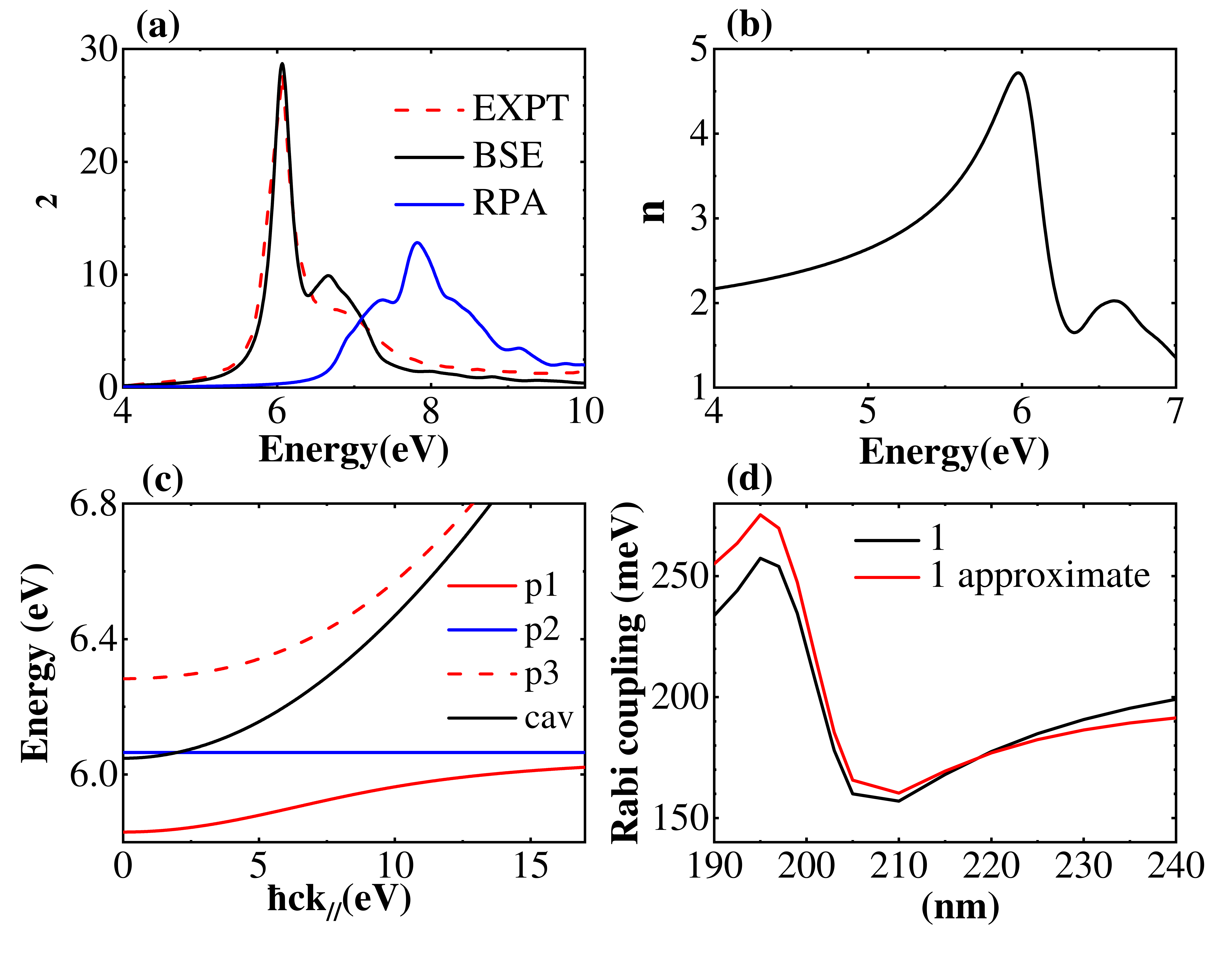}
		\caption{ \label{fig6}
			Exciton and polariton properties for all-dielectric hBN micrcavity.
			(a)Absorption spectrum of bulk hBN calculated by BSE (black solid line) and RPA (blue solid line) compared with the experimental results\cite{PhysRevB.40.7852} (red dashed line).
			(b) Refractive index calculated from the BSE reults.
			(c) Dispersion of the three branches of polaritons and cavity photon (black line).
			(d) Rabi coupling of polaritons composited by cavity photon mode and exciton 1(black line)  for different photon wavelength using Eq.~\ref{polariton2}.
			The corresponding results using approximate expression Eq.~\ref{approx} are in red line.
		}
\end{figure}

Now we investigate the all-dielectric hBN microcavity.
We first focus on the optical properties of bulk hBN.
The LDA ground state is obtained using a $6\times6\times2$ k-point mesh, along with a kinetic cutoff
of 110 Ry for the expansion of the wavefunctions.
The band structure is shown in Fig.~\ref{fig2}~(b).
The direct bandgap within LDA is 4.51~eV, located at $H$.
It is marked in the figure by dashed lines.
In BSE calculation we use a dense k-point mesh of $18\times18\times6$ and take the highest three valence bands and lowest two conduction bands as transition bands.
A scissor operator of 2.31 eV is applied to the Kohn-Sham energies.
The lowest exciton is mainly comprised by the transitons around $K$ with a binding energy of 0.76 eV.
The absorption spectrum is shown in Fig.~\ref{fig6}~(a).
We find that the RPA spectrum, which only takes inter-band absorption into account, deviates a lot from the experimental results.
As a comparision, the BSE spectrum with 0.17 eV broadening matches the experiment results much better, indicating that the excitonic effects is essensial here.
The refractive index calculated from BSE result by $n(\omega)=\sqrt{\frac{\text{Re}\epsilon+|\epsilon|}{2}}$
is shown in Fig.~\ref{fig6}~(b).
Near the absorption peak, the refractive index also shows a peak.
The degenerate exciton pair at 6.06 eV have the strongest oscillator strength, and we will only focus on them.
The structure of all-dielectric hBN microcavity is shown in Fig.~\ref{fig1}~(b).
The bulk hBN fills all the space of the microcavity between DBRs.
Using Eq.~\ref{polariton2} with $L^{\prime}=L$ and diagonalizing the following Hamiltonian:
\begin{equation}
		\hat{H}_{\mathrm{pol}}=
		\left(\begin{array}{ccc}
			E_{\text{cav}} & g_{1} & g_{2}\\
			g_{1} & E_{\text{exc1}} & 0\\
			g_{2} & 0 & E_{\text{exc2}}\\
		\end{array}\right),
\end{equation}
we get the exciton-polariton energy dispersion.
The results are shown in Fig.~\ref{fig6}~(c).
Two degenerate excitons together with photon mode form three exciton-polariton modes, labeled as p1, p2 and p3.
The middle branch, p2, is a linear combination of two excitons and the p1 and p3 branches are mixtures of two excitons and one photon modes.
The Rabi coupling is several times larger than the room temperature thermal energy at different photon wavelengths, as demonstrated in Fig.~\ref{fig6}~(d).
It minimum value at around 210 nm results from the maximum of the refractive index, which comes from the bright excitons at 6.06 eV.
In all-dielectric microcavity, there are more phonon modes scattering with the exciton-polaritons.
Intuitively, the dynamics toward BEC will be even easier.
However, the radiative lifetime of the excitons at finite temperature in bulk hBN is hard to be
estimated, as the exciton effective mass is ill-defined~\cite{paleari2019first}.
Consequently, a pure theoretical simulation of the BEC properties as that in monolayer is not
presented in this manuscript.
We note, however, that the radiative lifetime is approximately 800~ps at room temperature according to
a recent experiment~\cite{cao2013two}.
This value indicates that the decay of the exciton-polaritons in all-dielectric BN microcavity is even slower and the critical pumping strength will be low.
Therefore, bulk hBN based microcavity is also promising in applications related to exciton-polariton.
	
\subsection{Exciton-polariton mediated superconductivity}
\begin{figure}[h]
	\includegraphics[width=1.0\linewidth]{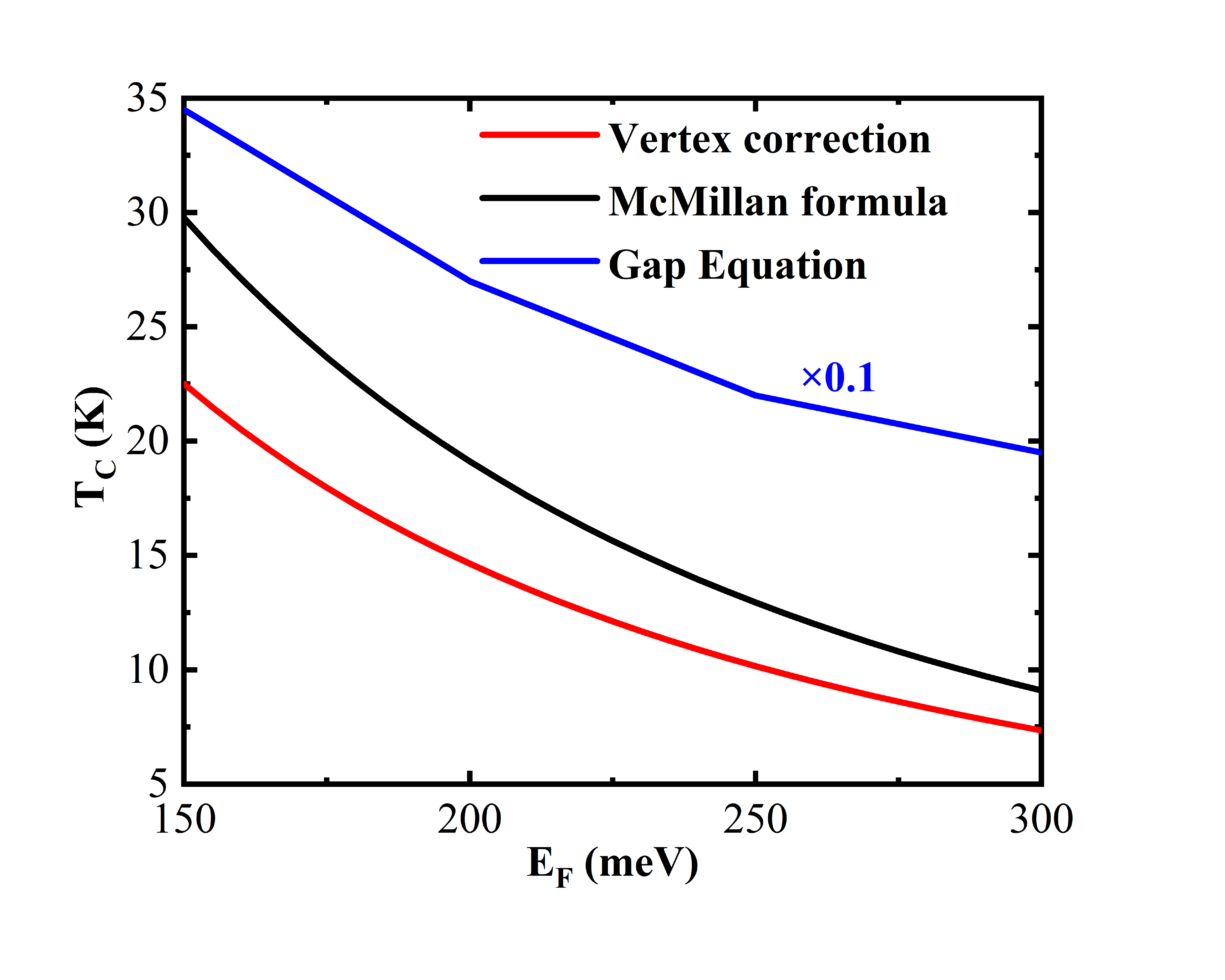}
	\caption{ \label{fig7}
		Exciton-polariton mediated superconductivity in BN based microcavity structure shown in Fig.~\ref{fig1}~(c),
		when $N_\text{c}=1\times10^{13}~$cm$^{-2}$,$\omega_\text{B}=150$meV, $L=2$~nm, $d=1$~nm, and $\epsilon=4$.
		The figure shows $T_\text{C}$ calculated by McMillan formula (Eq.~\ref{McMillan}, black line), by vertex correction formula (Eq.~\ref{Vertex}, red line) and by gap equation using Fr{\"o}hlich potential (Eqs.~\ref{fp}-\ref{delta}, blue line) with different 2DEG Fermi energy $E_\text{F}$.
	}
\end{figure}

Recently, several theoretical works have proposed that superconductivity can be induced by effective attractive interactions between the electrons mediated by exciton-polariton~\cite{PhysRevLett.104.106402, PhysRevLett.120.107001, sun2021theory}.
Here we investigate possibilities of this using hBN based microcavity.
The structure is shown in Fig.~\ref{fig1}~(c).
The dielectric constant of the vacuum space is chosen as 4.
We set the condensed exciton-polariton density to be a very large value, $N_\text{c}=1\times10^{13}~$cm$^{-2}$.
Now the characteristic bogolon energy $\omega_\text{B}$ is $\sim$150 meV.
Under an electric field perpendicular to the 2D layer, the condensated exciton-polaritons can interact with the electrons in the 2D electron gas QW.
We set the distance between these QWs as 2~nm.
These parameters are rather realistic.
For simplicity, we ignore the Coulomb interaction and electron-phonon coupling in the 2DEG to highlight the influence of exciton-polariton.
As shown in Fig.~\ref{fig7}, the superconducting transition temperature $T_\text{C}$ calculated by the gap equation using the Fr{\"o}hlich potential (Eqs.~\ref{fp}-\ref{delta}) is high, even up to the room temperature.
This is not physical because the large value of $T_\text{C}$ is a consequence of the Fr{\"o}hlich potential used~\cite{PhysRevB.93.054510} (Eq.~\ref{fp}).
Although the singularities of this potential can be removed by computing principal values numerically ~\cite{laussy2012superconductivity}, they will still develop two large shoulders in Eq.~\ref{EFF} ~\cite{PhysRevLett.104.106402}.
This may lead to a serious overestimation of $T_\text{C}$ .
In order to overcome these defects of the Fr{\"o}hlich potential, we apply the McMillan formula ~\cite{PhysRev.167.331, PhysRevB.12.905}
\begin{equation}\label{McMillan}
	T_\text{C}=\frac{\omega_\text{B}}{1.2} \exp \left[-\frac{1.04(1+\lambda)}{\lambda}\right],
\end{equation}
where $\lambda=-U_0(0)$ is the electron-bogolon coupling strength.
As shown in Fig.~\ref{fig7}, $T_\text{C}$ is substantially reduced, to several tens of Kelvin.
Upon this, we also note that the McMillan formula is only applicable when the the adiabatic approximation is valid and the Migdal therom holds~\cite{migdal1958interaction}, \textit{i}.\textit{e}. $\omega_\text{B} \ll E_\text{F}$.
Here $M\equiv\omega_\text{B}/E_\text{F} \in [0.5, 1]$, and the adiabatic approximation and Migdal therom  may fail.
To check how this impacts on the final results, we further consider the vertex correction and calculated $T_\text{C}$ by~\cite{grimaldi1995nonadiabatic}:
\begin{equation}\label{Vertex}
	T_\text{C}=\frac{1.13 \omega_\text{B}}{\sqrt{e}(1+M)} \exp \left[\frac{1}{2} \frac{M}{(1+M)}\right] \exp \left[-\frac{1+\lambda_{z}[1 /(1+M)]}{\lambda_{\Delta}}\right].
\end{equation}
The effective coupling $\lambda_{z}$ and $\lambda_{\Delta}$ are related to adiabatic coupling strength $\lambda$, and they are defined in Ref.~\onlinecite{grimaldi1995nonadiabatic} in details.
From Fig.~\ref{fig7}, we see that the vertex correction further reduces $T_\text{C}$ compared to the McMillan formula.
As $E_\text{F}$ gets larger, their $T_\text{C}$s tend to converge.
When the $E_\text{F}$ is smaller than $\omega_\text{B}$, the above two formulas are not applicable, and the $T_\text{C}$ should be calculated by the non-adiabatic limit fomula\cite{sadovskii2020antiadiabatic}:
\begin{equation}\label{NA}
	T_{\mathrm{c}} \sim \frac{E_\text{F}}{1+\frac{E_\text{F}}{\omega_{B}}} \exp \left(-\frac{1}{\lambda_{\text{NA}}}\right).
\end{equation}
Here $\lambda_{\text{NA}}=\frac{N(0)}{\pi} \int_{0}^{2 \pi} d \theta \frac{\left|M_{q}\right|^{2}}{\omega_{q}}$ and $q^{2}=2 m\left(\omega_{q}+E_{\mathrm{F}}\right)+k_{\mathrm{F}}^{2}-2 \sqrt{2 m\left(\omega_{q}+E_{\mathrm{F}}\right)} k_{\mathrm{F}} \cos \theta$.
$T_\text{C}$ is strongly suppressed and the value computed by \ref{NA} is vanishingly small.
We note, however, that the situation in the system considered belongs to the region when vertex correction 
is applicable ($M\equiv\omega_\text{B}/E_\text{F} \in [0.5, 1]$). 
Therefore, we conclude that $T_\text{C}$ can reach several tens of Kelvin.
Such 2D material based microcavities is promising for the fabrication of superconducting
devices, based on this scenario of exciton-polariton mediated superconductivity.

\section{CONCLUSION}
In summary, we studied the exciton-polariton properties in hBN based microcavities.
Using \textit{ab initio} calculations (DFT+BSE), we obtain exciton properties including absorption spectrum, exciton energy, exciton radiative lifetimes for hBN.
Based on these \textit{ab initio} results, we investigate the exciton-polariton dispersion, Rabi coupling, Hopfield coefficients, and exciton-polariton lifetimes for hBN based microcavities.
The oscillator strength in both monolayer and bulk hBN is very large, leading to large Rabi coupling.
There are five non-negligible exciton-polariton modes in the monolayer hBN and three in the bulk hBN.
We analyzed the component fraction and lifetime of each mode.
With the help of the Boltzmann equation, we find that room temperature exciton-polariton BEC can be
achieved in hBN-based microcavity owing to the large oscillator strength, binding energy, and the strong polariton-LO phonon interactions.
Superconductivity at a few tens of Kelvin may also be induced by polariton-electron interaction, if the microcavity structure is specially designed.
Overall, we conclude that hBN microcavities are very suitable platforms for studying the rich physics associated with exciton-polaritons.
And we hope this work can stimulate more experimental/theoretical studies and the predictions
presented in this manuscript can be further tested.
\begin{acknowledgments}
We acknowledge helpful discussions with Junren Shi and Shiwu Gao.
The authors are supported by the Beijing Natural Science Foundation under Grant No. Z200004,
the Strategic Priority Research Program of the Chinese Academy of Sciences Grant No. XDB33010400,
the National Basic Research Programs of China under Grand Nos. 2016YFA0300900, the National
Science Foundation of China under Grant Nos 11774003, 11934003, and 11634001.
The computational resources were provided by the supercomputer center in Peking University, China.
\end{acknowledgments}
		
%\bibliography{ref}

\end{document}